\documentclass[aps, pra, showpacs, twocolumn, amsfonts, amsmath, amssymb, superscriptaddress, floatfix]{revtex4-1}

\usepackage[T1]{fontenc}
\usepackage[latin9]{inputenc}
\usepackage{color}
\usepackage{graphicx}
\usepackage{ulem}
\usepackage{filemod}

\def\be{\begin{equation}}
\def\ee{\end{equation}}
\def\bea{\begin{eqnarray}}
\def\eea{\end{eqnarray}}

\begin{document}

\title{Anderson localization in the time domain}

\author{Krzysztof Sacha} 
\affiliation{
Instytut Fizyki imienia Mariana Smoluchowskiego, 
Uniwersytet Jagiello\'nski, ulica Profesora Stanis\l{}awa \L{}ojasiewicza 11, PL-30-348 Krak\'ow, Poland}
\affiliation{Mark Kac Complex Systems Research Center, Uniwersytet Jagiello\'nski, ulica Profesora Stanis\l{}awa \L{}ojasiewicza 11, PL-30-348 Krak\'ow, Poland
}

\author{Dominique Delande}
\affiliation{Laboratoire Kastler Brossel, UPMC-Sorbonne Universit\'es, CNRS, ENS-PSL Research University, Coll\`ege de France, 4 Place Jussieu, 75005 Paris, France}


\begin{abstract}
In analogy with usual Anderson localization taking place in time-independent disordered quantum systems where the disorder acts in configuration space, systems exposed to temporally disordered potentials can display Anderson localization in the time domain. We demonstrate this phenomenon with one-dimensional examples where a temporally disordered potential induces localization during the quantum evolution of wave-packets, in contrast with a fully delocalized classical dynamics. This is an example of a time crystal phenomenon, i.e., a crystalline behavior in the time domain. 
\end{abstract}


\maketitle

\section{Introduction}

Anderson localization (AL) is the inhibition of transport in a spatially disordered quantum system due to destructive interference \cite{Anderson1958}.
It manifests itself in the exponential localization
of eigenstates in configuration space, while the classical dynamics is on the average diffusive.
Relying on the interference between paths multiply scattered by the disorder -- e.g. a disordered
potential --  AL very much depends on the geometrical properties of these paths
and especially on the dimension of the system. The scaling theory of localization~\cite{Abrahams:Scaling:PRL79} makes it possible to understand semi-quantitatively this behavior and shows that AL is a generic
behavior in one-dimensional (1D) systems and in time-reversal invariant 2D systems. In higher dimensions,
localization usually takes place at low energy, while high energy states are delocalized. 

Although space and time do not play the same roles, one may wonder whether
a system where the potential is a spatially ``ordered'' function, but a temporally ``disordered'' function, displays
a phenomenon analogous to  AL, but in the time domain. It is the aim of this paper to show that the answer is positive.

A simple example of such a transposition between space and time is already known in the context of
localization phenomena. The so-called ``kicked rotor'' model describes a free rotor which is periodically
kicked by a spatially dependent potential. At long time, it displays ''dynamical localization'', that is
localization in momentum space~\cite{Moore:AtomOpticsRealizationQKR:PRL95}, which has been shown to be equivalent to AL~\cite{Fishman:LocDynAnders:PRL82}. Moreover, the addition of
temporal modulations of the kick strength is tantamount to adding spatial dimensions
in the equivalent Anderson model~\cite{Casati:IncommFreqsQKR:PRL89,Lemarie:Anderson3D:PRA09,Manai:Anderson2DKR:PRL15,Frahm:TypicalMap:PRE09}.

In the kicked rotor, the disordered character comes from the classically chaotic dynamics when the
kick strength is sufficiently strong. We will here use a different idea, namely directly 
incorporate the disordered character in the temporal variations of the potential.

Recently, it has been proposed that crystalline phenomena in the time domain can emerge due to spontaneous breaking of continuous time translation symmetry to a discrete one \cite{Wilczek2012,Li2012}. Such phenomena are termed time crystals and there is debate in the literature as to whether the formation of time crystals is possible \cite{Chernodub2013,Wilczek2013,Bruno2013,Wilczek2013a,Bruno2013a,Li2012a,Bruno2013b,Watanabe2015,Sacha2015,Sacha2015a,else16}. In solid state physics, it is often assumed that space crystals are already formed in a spontaneous process of space translation symmetry breaking and their properties are investigated with the help of space periodic potentials. A similar approach can be applied in the time domain. That is, periodically driven systems are able to reveal crystalline behavior in time in analogy to Bloch wave solutions of space periodic problems \cite{Sacha2015a}.
In this paper, we discuss a slightly related, but much more general, phenomenon. Using a conveniently tailored periodic
driving of a perfectly ordered system, one can create an effective disorder in the time domain. This may induce 
AL in the time domain, like extended Bloch states localize in the presence of a weak spatially disordered potential. The trick is to play with the harmonics of the driving frequency which are made resonant with the harmonics of the 
unperturbed periodic motion. It is the richness of the many-component interaction which allows us to control the localization properties. The phenomenon is at the same time robust and simple, since we can give an explicit recipe for the temporal driving.

\section{Results}
\subsection{Particle in a temporally disordered potential}

Let us begin with a time-independent system perturbed by a temporally disordered potential. 
A spatially constant potential 
trivially does not affect the dynamics. We have thus to use a space-dependent potential. 
To keep the spatial structure as simple as possible, we take a 1D particle
moving on a ring. The position of the particle is denoted by an angle $\theta$ and its momentum by $p$. The particle experiences a time dependent perturbation so that the classical Hamiltonian reads
\be
H=\frac{(p-\alpha)^2}{2}+Vg(\theta)f(t), 
\label{h}
\ee
where $V$ is the amplitude of the perturbation and the parameter $\alpha$ is introduced in order to remove degeneracy of unperturbed motions corresponding to $\pm p$. 
We assume that
\be
f(t+2\pi/\omega)=f(t)=\sum\limits_{k=-\infty}^{+\infty}f_ke^{ik\omega t},
\ee 
but, between $t$ and $t+2\pi/\omega,$ 
it performs random fluctuations, i.e. $f_k=f_{-k}^*$ are independent random variables. The function $g(\theta)$ is a regular function displaying no disorder. In the following, we concentrate on an example where $g(\theta)=\theta/\pi$ for $\theta\in[-\pi,\pi[$, i.e.
\be
g(\theta)=\sum\limits_{n=-\infty}^{+\infty}g_ne^{in\theta},
\ee 
where $g_n\!=\!\frac{i(-1)^{n}}{\pi n}$ for $n\ne 0$ and $g_0\!=\!0$. 

Switching to the rotating frame, $\Theta=\theta-\omega t$, we can see that $\Theta$ becomes a slowly varying quantity if its canonically conjugate momentum fulfills the resonance condition $P=p-\alpha-\omega\approx 0$. Then, the motion of the particle can be described by an effective Hamiltonian that is obtained by averaging the original Hamiltonian over the {\it fast time variable} (the so-called secular approximation \cite{Buchleitner2002}). The details of the calculation,
as well as tests of its validity, are given in the appendix. The time-averaged effective Hamiltonian writes:
\be
H_{\mathrm{eff}}=\frac{P^2}{2}+V\sum\limits_{k=-\infty}^{+\infty}g_k f_{-k} e^{ik\Theta}+\frac{\omega^2}{2}.
\label{heff}
\ee
For each realization of the random function $f(t)$, the motion of the particle in the vicinity of the resonant trajectory, i.e. for $|P|\ll \omega$, is described by an integrable Hamiltonian (\ref{heff}). The Hamiltonian (\ref{heff}) does not depend on $\omega$ apart from the last constant term. However, the validity of the effective Hamiltonian does depend on $\omega,$ as discussed in the appendix. The second order corrections to the effective Hamiltonian scale like $\frac{V^2}{\omega^2}$ and if this parameter goes to zero, the resonant motion of the system is perfectly described by (\ref{heff}).

The Hamiltonian (\ref{heff}) indicates that the particle effectively experiences a disordered potential. We will consider an example where 
\be
|g_kf_k|=\frac{1}{\sqrt{k_0}\pi^{1/4}}e^{-k^2/(2k_0^2)},
\label{gkfk}
\ee 
and Arg$(f_k)$ are random variables chosen uniformly in the interval $[0,2\pi[$. 
This requires that many $g_k$ coefficients are non-zero, that is a spatial dependance of the
potential with many Fourier components; it excludes potentials with a simple sinusoidal dependance. 
Actually, the secular term in the effective Hamiltonian (\ref{heff}) can be thought
of as the \emph{coherent} addition of resonant terms between the spatial harmonics of the
potential and the corresponding temporal harmonics of the disordered driving amplitude.

As we deal with a 1D system, AL is expected for any disorder strength. In our finite system with periodic boundary conditions, it is visible only if the localization length $\xi_{\mathrm{loc}}\ll 2\pi.$
In 1D systems, $\xi_{\mathrm{loc}}$ is essentially twice the transport mean free path~\cite{MuellerDelande:Houches:2009}. In the weak disorder limit, it can be computed
using the Born approximation, from the power spectrum of the potential correlation function and the disorder-free
density of states~\cite{MuellerDelande:Houches:2009,Kuhn:Speckle:NJP07}. In our case, this gives:
\begin{equation}
 \xi_{\mathrm{loc}} = \frac{k_0 \tilde E}{\sqrt{\pi} V^2}\ \exp\left(\frac{8\tilde E}{k_0^2}\right),
 \label{eq:xiloc}
\end{equation}
where $\tilde E=E-\omega^2/2$ is the energy of the eigenstate with respect to the average value of the potential energy in (\ref{heff}).
The exponential term is due to the finite correlation length $\zeta=\frac{\sqrt{2}}{k_0}$ of the disordered potential.

The Born approximation is valid in the weak scattering
regime, that is if $V^2\ll \tilde EE_{\zeta}$ where
$E_{\zeta}=1/\zeta^2=k_0^2/2$ is the so-called correlation energy \cite{Kuhn:Speckle:NJP07} (throughout this paper,
we assume $\hbar\!=\!1$). $E_{\zeta}$ is the energy scale separating two qualitatively
different regimes. If $\tilde E\ll E_{\zeta},$ the de Broglie wavelength of the particle is longer than $\zeta$, meaning that the particle can tunnel through potential hills; this regime is
often referred to as the ``quantum'' regime. In contrast, $\tilde E\gg E_{\zeta}$ refers to the ``semiclassical'' regime.
We are mainly interested in the ``quantum'' regime, where the exponential factor in eq.~(\ref{eq:xiloc}) is close to unity,
so that 
\be 
\frac{\xi_{\mathrm{loc}}}{\zeta} \approx \sqrt{\frac{2}{\pi}} \frac{E_\zeta \tilde E}{V^2}.
\ee 
Therefore, the weak scattering condition reduces to $\xi_{\mathrm{loc}}\gg \zeta.$ 

\begin{figure}
\includegraphics[width=0.9\columnwidth]{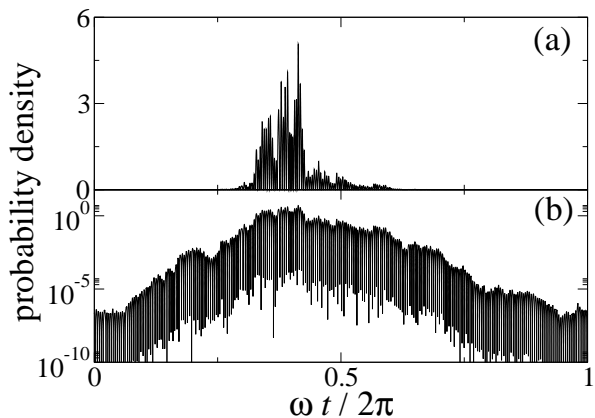}
\caption{Vizualization of an eigenstate of the Hamiltonian (\ref{heff}) in the laboratory frame. Panels show probability density for the measurement of a particle at a fixed point on a ring versus time, for an eigenstate of the Hamiltonian (\ref{heff}) of
energy $E-\frac{\omega^2}{2}=8\times 10^3$ for $k_0=10^3$, $V=4\times 10^3$. The localization length is of the order of $0.17/\omega$. The upper plot is on a linear scale, the lower plot on a logarithmic scale, showing approximate exponential localization.}
\label{fig1}
\end{figure}

In Fig.~\ref{fig1}, an exemplary eigenstate of the quantum version of the effective Hamiltonian (\ref{heff}) is presented for $V\!=\!4\times 10^3$, $k_0\!=\!10^3$ and $\alpha\!=\!(\sqrt{5}-1)/2$ 
\footnote{The role of the parameter $\alpha$ is to remove the degeneracy between the two counter-propagating waves at $p=\pm \omega$ which are simultaneously resonant. 
If $\alpha=0$ one would deal with (however very small) tunneling effects between counter propagating particles and the true eigenstates would be linear combinations of the states considered here and their counter-propagating mirror images.}.
The energy of this state is close to $\tilde E=E-\omega^2/2\!=\!8\times 10^3,$ so that $V^2/\tilde EE_{\zeta}=0.004\ll 1,$
and the Born approximation can be used. It predicts $\xi_{\mathrm{loc}}\!=\!0.30.$ Numerical calculation using the transfer matrix method~\cite{Lugan:Anderson1D:PRA09} agrees with this value within 1\%.
Single eigenstates do not display a perfect exponential localization, so that there are state-to-state fluctuations
of the estimated localization length, making the observed length 0.17 fully compatible with the theoretical calculation.

The original Hamiltonian (\ref{h}) is a periodic function of time. Thus, in the quantum description we can employ the Floquet theorem \cite{Buchleitner2002}, which is the analogue of the Bloch theorem in the time domain, and calculate quasi-energy eigenstates. These eigenstates, called Floquet states, are periodic eigenfunctions of the Floquet Hamiltonian $H_F=H(t)-i\partial_t$. 
Eigenstates of the effective Hamiltonian (\ref{heff}) are directly related to Floquet states within the secular approximation. When we return from the rotating frame to the laboratory frame, all eigenstates of (\ref{heff}) perform a periodic motion with period $\frac{2\pi}{\omega}$. Anderson localized eigenstates of (\ref{heff}) appear in the laboratory frame as periodically evolving exponentially localized density profiles. If we fix the position in the laboratory frame, we observe that probability density for a measurement of the particle at this position reveals a profile that is exponentially localized around a certain $t_0$, see Fig.~\ref{fig1}. The profile comes back periodically with period $2\pi/\omega$. Thus, one deals with the situation analogue to a time-independent 1D problem of a particle in a disordered potential with periodic boundary conditions.

AL considered here is induced by a time fluctuating perturbation not by a spatial disorder and appears in the time domain. In the case of a particle on a ring, an exponential localization is observed in time (for a fixed position in the configuration space) but also in the configuration space (at a fixed time) which is a specific property of the ring problem. In general, an exponential localization is only present in the time domain, as discussed in Sec.~\ref{gdis}.

\subsection{Model of time crystals with temporal disorder}

\begin{figure}
\includegraphics[width=0.9\columnwidth]{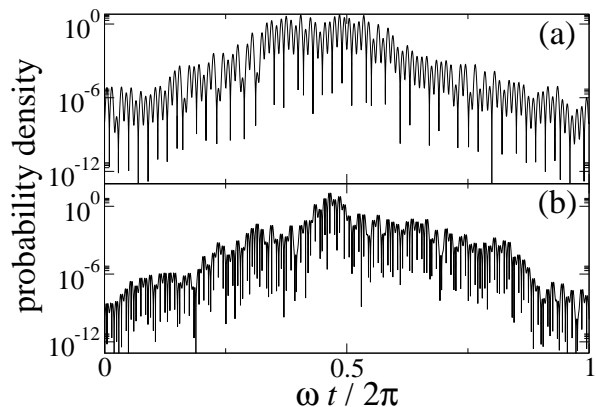}
\caption{Vizualization of eigenstates of the Hamiltonian (\ref{heff1}) in the laboratory frame. Panels show probability density for the measurement of a particle at a fixed point on a ring versus time, for eigenstates with energy in the middle of the lowest (a) and first excited (b) energy bands of the system described by the Hamiltonian (\ref{heff1}) for $s=100$, $\lambda=2\times 10^4$, $k_0=100$, $V=10$ (a) and $V=300$ (b). The localization lengths are both of the order of $0.16/\omega$.}
\label{fig2}
\end{figure}

We now consider a slightly different problem. The formation of time crystals is related to spontaneous breaking of time translation symmetry, in analogy with the spontaneous breaking of space translation symmetry in the formation of usual space crystals \cite{Wilczek2012,Li2012}. We will not consider the formation of time crystals, but try to simulate time crystal phenomena with the help of periodically driven systems \cite{Sacha2015a}. This is in analogy with the standard approach of solid state physics where space periodic Hamiltonians are used to model properties of space crystals. Let us assume that an unperturbed Hamiltonian for a particle on a ring reads
\be
{\cal H}_0=\frac{(p-\alpha)^2}{2}+\lambda\cos(s\theta)\cos(s\omega t), 
\ee 
where the integer $s\gg 1$. When we switch to the rotating frame, the new variable $\Theta=\theta-\omega t$ is a slow variable (similarly to the previous discussion) and the secular effective Hamiltonian 
\be
{\cal H}_{0,\mathrm{eff}}=\frac{P^2}{2}+\frac{\lambda}{2}\cos(s\Theta)+\frac{\omega^2}{2},
\ee 
describes quantitatively the resonant motion if ${\lambda^2/\omega^2} \ll 1$. Our problem has been reduced to a particle in an external periodic potential. 
The eigenstates of ${\cal H}_{0,\mathrm{eff}}$ are solutions of the Mathieu equation \cite{Buchleitner2002}.
In the limit of large amplitude $\lambda$ and large $s$ (with $\sqrt{\lambda}/s\gg 1$), they
appear in the laboratory frame as trains of $s$ localized wavepackets arriving periodically at any chosen point on the ring. This is a model of time crystals obtained with the help of a periodically driven system~\cite{Sacha2015a}. If we add to the unperturbed Hamiltonian ${\cal H}_0$ the same perturbation than in (\ref{h}), the full effective Hamiltonian of the system reads,
\be
{\cal H}_{\mathrm{eff}}=\frac{P^2}{2}+\frac{\lambda}{2}\cos(s\Theta)+V\sum\limits_{k=-\infty}^{+\infty}g_k f_{-k} e^{ik\Theta}+\frac{\omega^2}{2}.
\label{heff1}
\ee
For $V\ll\lambda$, we deal with a particle in a periodic lattice potential which has $s$  minima at
positions $\Theta=(2j+1)\pi/s$ for $0\leq j < s,$ perturbed by a weak disordered potential. If we can restrict to the lowest energy band of the periodic lattice, we can further simplify the description by deriving the standard Anderson Hamiltonian \cite{Sacha2015a}.
When $\zeta$ is shorter than  $2\pi/s$, the disorder is uncorrelated
between consecutive sites and the only parameters of the model are the disorder strength $V$ and the hopping amplitude $J$ between consecutive sites.
The latter has been computed in ~\cite{Zwerger:OpticalLattice:JPB03} in the limit of a deep lattice and is given by $J=(2^{5} \lambda^{3}s^2/\pi^{2})^{1/4} \exp( - \sqrt{32\lambda}/s)$.
In the weak disorder limit $V\ll J,$ the localization length at the center of the band reads $\xi_{\mathrm{loc}}=8\pi J^2/sV^2$~\cite{MuellerDelande:Houches:2009}.
Figure~\ref{fig2}(a) shows the evolution of the probability density for the particle detection at a fixed position in the laboratory frame, for an  eigenstate in the middle of the band.
For the parameters of the figure, we obtain $J=7.57,$ so that we are not fully in the perturbative limit.
Nevertheless the predicted localization length is $\xi_{\mathrm{loc}}=0.144$ in good agreement with 
the results in the figure.
Note that the Hamiltonian (\ref{heff1}) also predicts the possibility of AL in higher energy bands.
In the first excited band, the asymptotic solutions of the Mathieu equation~\cite{AbramovitzStegun:64} predict
$J$ to be 32 times larger than in the lowest band. For a $V$ value 30 times larger than in Fig.~\ref{fig2}(a), the localization length should be similar to the case of the lowest band; that is indeed observed in Fig.~\ref{fig2}(b).  

\subsection{General discussion}
\label{gdis}

We have illustrated AL in the time domain with the simple problem of a particle on a ring. However, the phenomenon is general and can be realized in a broad class of systems. Assume that we have an integrable 1D system described by a classical Hamiltonian $H_0(x,p)$ which is perturbed by a time-dependent potential of the form $Vg(x)f(t)$ where $f(t)$ is a randomly fluctuating function but fulfills $f(t+2\pi/\omega)=f(t)$ and $g(x)$ is a regular function. For example, we can think of a particle bouncing on a mirror whose position fluctuates in time; then, $H_0(x,p)=\frac{p^2}{2}+x$ and $g(x)=x$ \cite{Buchleitner2002}. The canonical transformation to the angle-action variables $(\theta,J)$ of the unperturbed problem allows one to write a full Hamiltonian in the form $H=H_0(J)+Vg(\theta,J)f(t)$. Periodic solutions of the unperturbed system possess very simple forms in the angle-action variables, i.e. $J=$constant and $\theta=H'_0(J)t+$constant, where $H'_0=\frac{\partial H_0}{\partial J}$. Let us stress that the canonical transformation between $(x,p)$ and $(\theta,J)$ is usually non-linear. On the other hand, the previous equation shows that the angle variable $\theta$ changes always linearly with time on an unperturbed orbit.
By expanding $g(J,\theta)=\sum_ng_n(J)e^{in\theta}$ and $f(t)$ in Fourier series, 
one can perform the standard approach for resonant motion that yields the effective secular Hamiltonian 
\be
H_{\mathrm{eff}}=\frac12 H_0''(J_0)(J-J_0)^2+V\sum\limits_{k=-\infty}^{+\infty}g_k(J_0)f_{-k}e^{ik\Theta}, 
\label{heff2}
\ee
modulo a constant term, where $\Theta=\theta-\omega t$ and $J_0$ fulfills the resonant condition $\omega=H_0'(J_0)$. Such an effective Hamiltonian has the same form than (\ref{heff}). If, in addition, we apply a periodic driving $\lambda g(x)\cos(s\omega t)$, the resulting effective Hamiltonian will be of the form (\ref{heff1})~\footnote{In general, there can be several resonances between the external
driving at frequency $s\omega$ and the internal frequency $\partial H_0/\partial J.$ We assume $\lambda$ to be small enough for these resonances not to overlap.}. Thus, in terms of the angle-action variables, we obtain the same kind of behavior: AL phenomena can be expected. There is, however, one difference. AL is identified and described in the rotating frame in terms of angle-action variables. When one tries to plot the probability density of a localized state in the laboratory frame as a function of $x$ (for fixed $t$), one does not observe in general an exponentially localized profile because the canonical transformation from $(\Theta,J)$ to $(x,p)$ is generically non-linear. However, the temporal evolution of the probability of detecting the particle at a fixed position $x$ is always exponentially localized around a certain $t_0$. Indeed, our temporal AL takes place along a periodic orbit of the unperturbed system, resonant with the time-dependent perturbation. As the relation between $t$ and $\Theta$ is always linear for an unperturbed motion, AL in the $\Theta$ space also means exponential localization in time.

Finally let us discuss possible realizations of AL in the time domain in higher dimensional systems. Assume that, in a 3D case, the Hamiltonian of a one-particle system can be written, in the angle-action variables $(\theta_i,J_i)$, as $H_0(J_1)+V(\theta_1,\theta_2,\theta_3;J_1,J_2,J_3;t)$ where $V$ fluctuates in time but fulfills $V(t+2\pi/\omega)=V(t)$. This occurs, e.g., in the Hydrogen atom perturbed by a fluctuating electromagnetic field where $J_1$ is the 
principal action, $\theta_1$ denotes the position of 
the electron on an unperturbed elliptical orbit, while $(\theta_2,J_2,\theta_3,J_3)$ define 
the shape and orientation of the orbit. It may happen that, even if the perturbation is on, there is a stable resonant periodic orbit where $(\theta_2,J_2,\theta_3,J_3)$ remain nearly constant \cite{Buchleitner2002}. Then, the effective Hamiltonian of the $\theta_1$ degree of freedom can be reduced to a form similar to (\ref{heff2}) and AL of the electron can take place.

\section{Summary}

To summarize, we have proposed a new scenario for the observation of Anderson localization, based on a 
temporally disordered driving of a quantum system. In proper conditions, it leads to the spontaneous
appearance of wavepackets Anderson localized along an unperturbed periodic trajectory.

A key point for a possible experimental observation of AL in the time domain is that
both the temporally ``disordered'' periodic driving and the ``regular'' inner potential must contain many harmonics of the fundamental frequency. Thanks to modern electronics, creating a complex temporal profile of the driving is not necessarily difficult. Getting a highly anharmonic potential requires a specific system.
In the example used in this paper, the high-order spatial harmonics come from the discontinuity of the potential $g(\theta)$ in (\ref{h}). In a more realistic system, it will require some kind of singularity. A first example could be
a cold atom bouncing on an optical mirror (created by an evanescent wave \cite{Szriftgiser:Mirror:PRL96}) whose effective position is modulated in time. Another possibility would be to use the dynamics of a Rydberg electron \cite{Maeda2004}. There the close encounters with the nucleus are responsible for a highly anharmonic classical motion
and consequently Fourier components decrease only slowly at large order~\cite{Buchleitner2002}. Driving the Rydberg electron with a temporally disordered microwave field could lead to AL in the time domain. Let us also mention the possibility of trapping a cold atomic gas in a ring-shaped trap whose parameters
can be modulated in time or the use of rings of superconducting or normal metal devices \cite{Bleszynski2009}.

\section*{Acknoledgments}

This work was performed within the Polish-French bilateral POLONIUM Grant 33162XA and the FOCUS action of Faculty of Physics, Astronomy and Applied Computer Science of Jagiellonian University. Support of EU Horizon-2020 QUIC 641122 FET program is also acknowledged. 

\section*{Appendix}

In this Appendix we analyze the validity of the effective Hamiltonian (\ref{heff}) obtained within the so-called secular approximation \cite{Buchleitner2002,Lichtenberg_s}.

The original Hamiltonian (\ref{h}) describes a particle on a ring.
Let us switch to the rotating frame with the help of the following canonical transformation
\begin{eqnarray}
\Theta & = & \theta -\omega t\\
P & = & p- \alpha -\omega. 
\end{eqnarray}
A straightforward algebraic manipulation shows that the Hamiltonian in the new coordinates is
\begin{equation}
H = \frac{P^2}{2} + V \sum_{m=-\infty}^{+\infty}{\left(\sum_{k=-\infty}^{+\infty}{g_kf_{m-k}\ e^{ik\Theta}}\right)\ e^{im\omega t}} +\frac{\omega^2}{2}.
\end{equation}

The secular approximation is valid at large $\omega.$ In the region where $|P|\ll \omega,$ the Hamilton's equations of motion show
that the evolution is slow for the $\Theta,P$ variables, while $\omega t$ is a rapidly varying variable. This makes it possible at first order to
keep in the Hamiltonian only the secular terms $m=0,$ resulting in the effective Hamiltonian (\ref{heff}).

\begin{figure}
\includegraphics[width=1.\columnwidth]{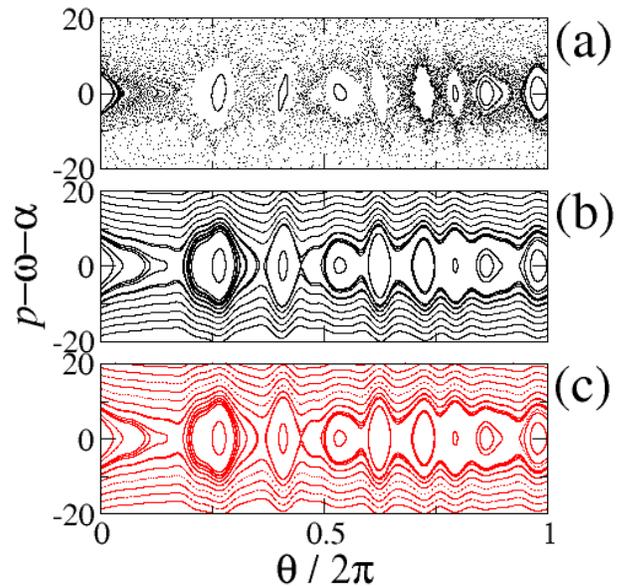}
\caption{(color online) Poincar\'e surface of section for a particle moving on a ring in the presence of a temporally disordered perturbation for $V=20$, $k_0=10$ and $\alpha=(\sqrt{5}-1)/2$. (a) and (b) correspond to the exact results for $\omega=300-\alpha$ and $\omega=2000-\alpha$, respectively. (c) is related to the phase space portrait generated by the effective Hamiltonian (\ref{heff}).}
\label{sos_s}
\end{figure}

\begin{figure}
\includegraphics[width=1.\columnwidth]{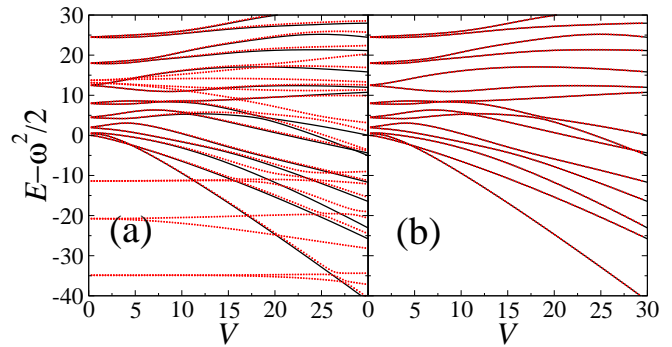}
\caption{(color online) Black lines show a bunch of low lying energy levels of the effective Hamiltonian (\ref{heff}) versus $V$ for $k_0=10$. Red circles correspond to quasienergy levels of the full Hamiltonian (\ref{h}) for $\omega=300-\alpha$ (a) and $\omega=2000-\alpha$ (b) where $\alpha=(\sqrt{5}-1)/2$. In (b) red circles follow closely black lines and the latter are hardly visible.}
\label{level_s}
\end{figure}

\begin{figure}
\includegraphics[width=1.\columnwidth]{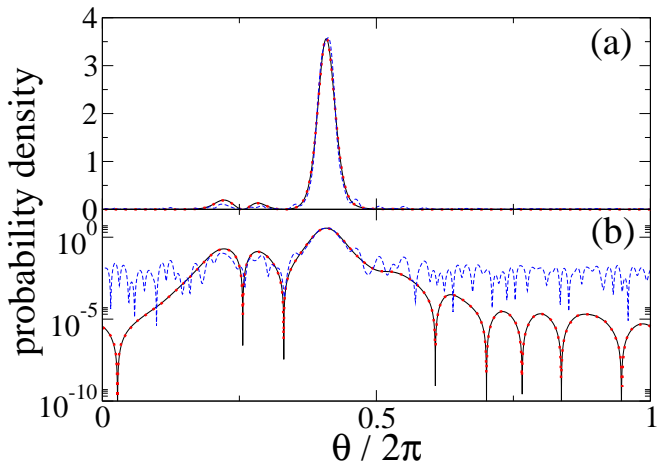}
\caption{(color online) Black solid lines show the probability density of the $7^{\mathrm{th}}$ 
excited eigenstate of the effective Hamiltonian (\ref{heff}) in the laboratory frame in linear (a) and logarithmic (b) scale. Blue dash lines depict the probability density of the corresponding Floquet eigenstate of the full Hamiltonian (\ref{h}), plotted for a fixed moment of time, for $\omega=300-\alpha$. Red dotted lines (hardly distinguishable from the black solid lines) show results similar to the blue dash lines but for $\omega=2000-\alpha$. The other parameters of the system are the following: $V=20$, $k_0=10$ and $\alpha=(\sqrt{5}-1)/2$.}
\label{eigens_s}
\end{figure}

The effect of non-secular terms $m\ne 0$ can be estimated at second order in $V/\omega.$ It results in the contribution to the effective Hamiltonian \cite{Buchleitner2002,Lichtenberg_s}:
\bea
H^{(2)}&=&-\frac{V^2}{2\omega^2}\sum_{m\ne 0}{\sum_{n,n'}{\frac{nn'g_ng_{n'}f_{-n+m}f_{-n'+m}\;e^{i(n+n')\Theta}}{m^2}}}. \cr &&
\eea
If $V^2/\omega^2\rightarrow 0$, the second order terms disappear. In that limit, we may expect that the effective Hamiltonian (\ref{heff}) reproduces the exact resonance dynamics of the system. It can be confirmed in numerical simulations where predictions based on the effective Hamiltonian are compared with the exact results.

In Fig.~\ref{sos_s} we compare Poincar\'e surface of sections, obtained in numerical integration of classical equations of motion generated by the exact Hamiltonian (\ref{h}), with the phase space portrait corresponding to the effective Hamiltonian (\ref{heff}). We can see that for sufficiently large $\omega$ the exact results follow precisely the secular approximation.

Switching to quantum description, we can compare eigenstates of the exact Floquet Hamiltonian $H_F=H-i\partial_t$ with the corresponding eigenstates of the effective Hamiltonian (\ref{heff}). Figure~\ref{level_s} presents quasienergy levels versus $V$ for two different values of $\omega$. Example of an eigenstate is shown in Fig.~\ref{eigens_s}. Similarly to the classical case, the effective Hamiltonian provides an accurate description of the system if $V^2/\omega^2$ is sufficiently small.

To sum up, the effective Hamiltonian (\ref{heff}) provides quantitative description of the resonant behavior of the system if $V^2/\omega^2$ goes to zero. We have illustrated the validity of such a secular approximation when the effective disordered potential in (\ref{heff}) is characterized by the correlation length $\zeta={\sqrt{2}/k_0}=0.14$, cf. (\ref{gkfk}). Diagonalization of the full Floquet Hamiltonian for much smaller $\zeta$ becomes very difficult numerically. However, thanks to the effective Hamiltonian approach, such a regime can be investigated.

\bibliographystyle{apsrev4-1}

\begin{thebibliography}{31}%
\makeatletter
\providecommand \@ifxundefined [1]{%
 \@ifx{#1\undefined}
}%
\providecommand \@ifnum [1]{%
 \ifnum #1\expandafter \@firstoftwo
 \else \expandafter \@secondoftwo
 \fi
}%
\providecommand \@ifx [1]{%
 \ifx #1\expandafter \@firstoftwo
 \else \expandafter \@secondoftwo
 \fi
}%
\providecommand \natexlab [1]{#1}%
\providecommand \enquote  [1]{``#1''}%
\providecommand \bibnamefont  [1]{#1}%
\providecommand \bibfnamefont [1]{#1}%
\providecommand \citenamefont [1]{#1}%
\providecommand \href@noop [0]{\@secondoftwo}%
\providecommand \href [0]{\begingroup \@sanitize@url \@href}%
\providecommand \@href[1]{\@@startlink{#1}\@@href}%
\providecommand \@@href[1]{\endgroup#1\@@endlink}%
\providecommand \@sanitize@url [0]{\catcode `\\12\catcode `\$12\catcode
  `\&12\catcode `\#12\catcode `\^12\catcode `\_12\catcode `\%12\relax}%
\providecommand \@@startlink[1]{}%
\providecommand \@@endlink[0]{}%
\providecommand \url  [0]{\begingroup\@sanitize@url \@url }%
\providecommand \@url [1]{\endgroup\@href {#1}{\urlprefix }}%
\providecommand \urlprefix  [0]{URL }%
\providecommand \Eprint [0]{\href }%
\providecommand \doibase [0]{http://dx.doi.org/}%
\providecommand \selectlanguage [0]{\@gobble}%
\providecommand \bibinfo  [0]{\@secondoftwo}%
\providecommand \bibfield  [0]{\@secondoftwo}%
\providecommand \translation [1]{[#1]}%
\providecommand \BibitemOpen [0]{}%
\providecommand \bibitemStop [0]{}%
\providecommand \bibitemNoStop [0]{.\EOS\space}%
\providecommand \EOS [0]{\spacefactor3000\relax}%
\providecommand \BibitemShut  [1]{\csname bibitem#1\endcsname}%
\let\auto@bib@innerbib\@empty
\bibitem [{\citenamefont {Anderson}(1958)}]{Anderson1958}%
  \BibitemOpen
  \bibfield  {author} {\bibinfo {author} {\bibfnamefont {P.~W.}\ \bibnamefont
  {Anderson}},\ }\href {\doibase 10.1103/PhysRev.109.1492} {\bibfield
  {journal} {\bibinfo  {journal} {Phys. Rev.}\ }\textbf {\bibinfo {volume}
  {109}},\ \bibinfo {pages} {1492} (\bibinfo {year} {1958})}\BibitemShut
  {NoStop}%
\bibitem [{\citenamefont {Abrahams}\ \emph {et~al.}(1979)\citenamefont
  {Abrahams}, \citenamefont {Anderson}, \citenamefont {Licciardello},\ and\
  \citenamefont {Ramakrishnan}}]{Abrahams:Scaling:PRL79}%
  \BibitemOpen
  \bibfield  {author} {\bibinfo {author} {\bibfnamefont {E.}~\bibnamefont
  {Abrahams}}, \bibinfo {author} {\bibfnamefont {P.~W.}\ \bibnamefont
  {Anderson}}, \bibinfo {author} {\bibfnamefont {D.~C.}\ \bibnamefont
  {Licciardello}}, \ and\ \bibinfo {author} {\bibfnamefont {T.~V.}\
  \bibnamefont {Ramakrishnan}},\ }\href {\doibase 10.1103/PhysRevLett.42.673}
  {\bibfield  {journal} {\bibinfo  {journal} {Phys. Rev. Lett.}\ }\textbf
  {\bibinfo {volume} {42}},\ \bibinfo {pages} {673} (\bibinfo {year}
  {1979})}\BibitemShut {NoStop}%
\bibitem [{\citenamefont {Moore}\ \emph {et~al.}(1995)\citenamefont {Moore},
  \citenamefont {Robinson}, \citenamefont {Bharucha}, \citenamefont
  {Sundaram},\ and\ \citenamefont
  {Raizen}}]{Moore:AtomOpticsRealizationQKR:PRL95}%
  \BibitemOpen
  \bibfield  {author} {\bibinfo {author} {\bibfnamefont {F.~L.}\ \bibnamefont
  {Moore}}, \bibinfo {author} {\bibfnamefont {J.~C.}\ \bibnamefont {Robinson}},
  \bibinfo {author} {\bibfnamefont {C.~F.}\ \bibnamefont {Bharucha}}, \bibinfo
  {author} {\bibfnamefont {B.}~\bibnamefont {Sundaram}}, \ and\ \bibinfo
  {author} {\bibfnamefont {M.~G.}\ \bibnamefont {Raizen}},\ }\href {\doibase
  10.1103/PhysRevLett.75.4598} {\bibfield  {journal} {\bibinfo  {journal}
  {Phys. Rev. Lett.}\ }\textbf {\bibinfo {volume} {75}},\ \bibinfo {pages}
  {4598} (\bibinfo {year} {1995})}\BibitemShut {NoStop}%
\bibitem [{\citenamefont {Fishman}\ \emph {et~al.}(1982)\citenamefont
  {Fishman}, \citenamefont {Grempel},\ and\ \citenamefont
  {Prange}}]{Fishman:LocDynAnders:PRL82}%
  \BibitemOpen
  \bibfield  {author} {\bibinfo {author} {\bibfnamefont {S.}~\bibnamefont
  {Fishman}}, \bibinfo {author} {\bibfnamefont {D.~R.}\ \bibnamefont
  {Grempel}}, \ and\ \bibinfo {author} {\bibfnamefont {R.~E.}\ \bibnamefont
  {Prange}},\ }\href {\doibase 10.1103/PhysRevLett.49.509} {\bibfield
  {journal} {\bibinfo  {journal} {Phys. Rev. Lett.}\ }\textbf {\bibinfo
  {volume} {49}},\ \bibinfo {pages} {509} (\bibinfo {year} {1982})}\BibitemShut
  {NoStop}%
\bibitem [{\citenamefont {Casati}\ \emph {et~al.}(1989)\citenamefont {Casati},
  \citenamefont {Guarneri},\ and\ \citenamefont
  {Shepelyansky}}]{Casati:IncommFreqsQKR:PRL89}%
  \BibitemOpen
  \bibfield  {author} {\bibinfo {author} {\bibfnamefont {G.}~\bibnamefont
  {Casati}}, \bibinfo {author} {\bibfnamefont {I.}~\bibnamefont {Guarneri}}, \
  and\ \bibinfo {author} {\bibfnamefont {D.~L.}\ \bibnamefont {Shepelyansky}},\
  }\href {\doibase 10.1103/PhysRevLett.62.345} {\bibfield  {journal} {\bibinfo
  {journal} {Phys. Rev. Lett.}\ }\textbf {\bibinfo {volume} {62}},\ \bibinfo
  {pages} {345} (\bibinfo {year} {1989})}\BibitemShut {NoStop}%
\bibitem [{\citenamefont {Lemari\'e}\ \emph {et~al.}(2009)\citenamefont
  {Lemari\'e}, \citenamefont {Chab\'e}, \citenamefont {Szriftgiser},
  \citenamefont {Garreau}, \citenamefont {Gr\'emaud},\ and\ \citenamefont
  {Delande}}]{Lemarie:Anderson3D:PRA09}%
  \BibitemOpen
  \bibfield  {author} {\bibinfo {author} {\bibfnamefont {G.}~\bibnamefont
  {Lemari\'e}}, \bibinfo {author} {\bibfnamefont {J.}~\bibnamefont {Chab\'e}},
  \bibinfo {author} {\bibfnamefont {P.}~\bibnamefont {Szriftgiser}}, \bibinfo
  {author} {\bibfnamefont {J.~C.}\ \bibnamefont {Garreau}}, \bibinfo {author}
  {\bibfnamefont {B.}~\bibnamefont {Gr\'emaud}}, \ and\ \bibinfo {author}
  {\bibfnamefont {D.}~\bibnamefont {Delande}},\ }\href {\doibase
  10.1103/PhysRevA.80.043626} {\bibfield  {journal} {\bibinfo  {journal} {Phys.
  Rev. A}\ }\textbf {\bibinfo {volume} {80}},\ \bibinfo {pages} {043626}
  (\bibinfo {year} {2009})},\ \Eprint {http://arxiv.org/abs/0907.3411}
  {arXiv:0907.3411 [quant-ph]} \BibitemShut {NoStop}%
\bibitem [{\citenamefont {Manai}\ \emph {et~al.}(2015)\citenamefont {Manai},
  \citenamefont {Cl{\'e}ment}, \citenamefont {Chicireanu}, \citenamefont
  {Hainaut}, \citenamefont {Garreau}, \citenamefont {Szriftgiser},\ and\
  \citenamefont {Delande}}]{Manai:Anderson2DKR:PRL15}%
  \BibitemOpen
  \bibfield  {author} {\bibinfo {author} {\bibfnamefont {I.}~\bibnamefont
  {Manai}}, \bibinfo {author} {\bibfnamefont {J.-F.}\ \bibnamefont
  {Cl{\'e}ment}}, \bibinfo {author} {\bibfnamefont {R.}~\bibnamefont
  {Chicireanu}}, \bibinfo {author} {\bibfnamefont {C.}~\bibnamefont {Hainaut}},
  \bibinfo {author} {\bibfnamefont {J.~C.}\ \bibnamefont {Garreau}}, \bibinfo
  {author} {\bibfnamefont {P.}~\bibnamefont {Szriftgiser}}, \ and\ \bibinfo
  {author} {\bibfnamefont {D.}~\bibnamefont {Delande}},\ }\href {\doibase
  10.1103/PhysRevLett.115.240603} {\bibfield  {journal} {\bibinfo  {journal}
  {Phys. Rev. Lett.}\ }\textbf {\bibinfo {volume} {115}},\ \bibinfo {pages}
  {240603} (\bibinfo {year} {2015})}\BibitemShut {NoStop}%
\bibitem [{\citenamefont {Frahm}\ and\ \citenamefont
  {Shepelyansky}(2009)}]{Frahm:TypicalMap:PRE09}%
  \BibitemOpen
  \bibfield  {author} {\bibinfo {author} {\bibfnamefont {K.~M.}\ \bibnamefont
  {Frahm}}\ and\ \bibinfo {author} {\bibfnamefont {D.~L.}\ \bibnamefont
  {Shepelyansky}},\ }\href {\doibase 10.1103/PhysRevE.80.016210} {\bibfield
  {journal} {\bibinfo  {journal} {Phys. Rev. E}\ }\textbf {\bibinfo {volume}
  {80}},\ \bibinfo {pages} {016210} (\bibinfo {year} {2009})}\BibitemShut
  {NoStop}%
\bibitem [{\citenamefont {Wilczek}(2012)}]{Wilczek2012}%
  \BibitemOpen
  \bibfield  {author} {\bibinfo {author} {\bibfnamefont {F.}~\bibnamefont
  {Wilczek}},\ }\href {\doibase 10.1103/PhysRevLett.109.160401} {\bibfield
  {journal} {\bibinfo  {journal} {Phys. Rev. Lett.}\ }\textbf {\bibinfo
  {volume} {109}},\ \bibinfo {pages} {160401} (\bibinfo {year}
  {2012})}\BibitemShut {NoStop}%
\bibitem [{\citenamefont {Li}\ \emph {et~al.}(2012{\natexlab{a}})\citenamefont
  {Li}, \citenamefont {Gong}, \citenamefont {Yin}, \citenamefont {Quan},
  \citenamefont {Yin}, \citenamefont {Zhang}, \citenamefont {Duan},\ and\
  \citenamefont {Zhang}}]{Li2012}%
  \BibitemOpen
  \bibfield  {author} {\bibinfo {author} {\bibfnamefont {T.}~\bibnamefont
  {Li}}, \bibinfo {author} {\bibfnamefont {Z.-X.}\ \bibnamefont {Gong}},
  \bibinfo {author} {\bibfnamefont {Z.-Q.}\ \bibnamefont {Yin}}, \bibinfo
  {author} {\bibfnamefont {H.~T.}\ \bibnamefont {Quan}}, \bibinfo {author}
  {\bibfnamefont {X.}~\bibnamefont {Yin}}, \bibinfo {author} {\bibfnamefont
  {P.}~\bibnamefont {Zhang}}, \bibinfo {author} {\bibfnamefont {L.-M.}\
  \bibnamefont {Duan}}, \ and\ \bibinfo {author} {\bibfnamefont
  {X.}~\bibnamefont {Zhang}},\ }\href {\doibase 10.1103/PhysRevLett.109.163001}
  {\bibfield  {journal} {\bibinfo  {journal} {Phys. Rev. Lett.}\ }\textbf
  {\bibinfo {volume} {109}},\ \bibinfo {pages} {163001} (\bibinfo {year}
  {2012}{\natexlab{a}})}\BibitemShut {NoStop}%
\bibitem [{\citenamefont {Chernodub}(2013)}]{Chernodub2013}%
  \BibitemOpen
  \bibfield  {author} {\bibinfo {author} {\bibfnamefont {M.~N.}\ \bibnamefont
  {Chernodub}},\ }\href {\doibase 10.1103/PhysRevD.87.025021} {\bibfield
  {journal} {\bibinfo  {journal} {Phys. Rev. D}\ }\textbf {\bibinfo {volume}
  {87}},\ \bibinfo {pages} {025021} (\bibinfo {year} {2013})}\BibitemShut
  {NoStop}%
\bibitem [{\citenamefont {Wilczek}(2013{\natexlab{a}})}]{Wilczek2013}%
  \BibitemOpen
  \bibfield  {author} {\bibinfo {author} {\bibfnamefont {F.}~\bibnamefont
  {Wilczek}},\ }\href {\doibase 10.1103/PhysRevLett.111.250402} {\bibfield
  {journal} {\bibinfo  {journal} {Phys. Rev. Lett.}\ }\textbf {\bibinfo
  {volume} {111}},\ \bibinfo {pages} {250402} (\bibinfo {year}
  {2013}{\natexlab{a}})}\BibitemShut {NoStop}%
\bibitem [{\citenamefont {Bruno}(2013{\natexlab{a}})}]{Bruno2013}%
  \BibitemOpen
  \bibfield  {author} {\bibinfo {author} {\bibfnamefont {P.}~\bibnamefont
  {Bruno}},\ }\href {\doibase 10.1103/PhysRevLett.110.118901} {\bibfield
  {journal} {\bibinfo  {journal} {Phys. Rev. Lett.}\ }\textbf {\bibinfo
  {volume} {110}},\ \bibinfo {pages} {118901} (\bibinfo {year}
  {2013}{\natexlab{a}})}\BibitemShut {NoStop}%
\bibitem [{\citenamefont {Wilczek}(2013{\natexlab{b}})}]{Wilczek2013a}%
  \BibitemOpen
  \bibfield  {author} {\bibinfo {author} {\bibfnamefont {F.}~\bibnamefont
  {Wilczek}},\ }\href {\doibase 10.1103/PhysRevLett.110.118902} {\bibfield
  {journal} {\bibinfo  {journal} {Phys. Rev. Lett.}\ }\textbf {\bibinfo
  {volume} {110}},\ \bibinfo {pages} {118902} (\bibinfo {year}
  {2013}{\natexlab{b}})}\BibitemShut {NoStop}%
\bibitem [{\citenamefont {Bruno}(2013{\natexlab{b}})}]{Bruno2013a}%
  \BibitemOpen
  \bibfield  {author} {\bibinfo {author} {\bibfnamefont {P.}~\bibnamefont
  {Bruno}},\ }\href {\doibase 10.1103/PhysRevLett.111.029301} {\bibfield
  {journal} {\bibinfo  {journal} {Phys. Rev. Lett.}\ }\textbf {\bibinfo
  {volume} {111}},\ \bibinfo {pages} {029301} (\bibinfo {year}
  {2013}{\natexlab{b}})}\BibitemShut {NoStop}%
\bibitem [{\citenamefont {Li}\ \emph {et~al.}(2012{\natexlab{b}})\citenamefont
  {Li}, \citenamefont {Gong}, \citenamefont {Yin}, \citenamefont {Quan},
  \citenamefont {Yin}, \citenamefont {Zhang}, \citenamefont {Duan},\ and\
  \citenamefont {Zhang}}]{Li2012a}%
  \BibitemOpen
  \bibfield  {author} {\bibinfo {author} {\bibfnamefont {T.}~\bibnamefont
  {Li}}, \bibinfo {author} {\bibfnamefont {Z.-X.}\ \bibnamefont {Gong}},
  \bibinfo {author} {\bibfnamefont {Z.-Q.}\ \bibnamefont {Yin}}, \bibinfo
  {author} {\bibfnamefont {H.~T.}\ \bibnamefont {Quan}}, \bibinfo {author}
  {\bibfnamefont {X.}~\bibnamefont {Yin}}, \bibinfo {author} {\bibfnamefont
  {P.}~\bibnamefont {Zhang}}, \bibinfo {author} {\bibfnamefont {L.-M.}\
  \bibnamefont {Duan}}, \ and\ \bibinfo {author} {\bibfnamefont
  {X.}~\bibnamefont {Zhang}},\ }\href@noop {} {\  (\bibinfo {year}
  {2012}{\natexlab{b}})},\ \Eprint {http://arxiv.org/abs/1212.6959}
  {arXiv:1212.6959} \BibitemShut {NoStop}%
\bibitem [{\citenamefont {Bruno}(2013{\natexlab{c}})}]{Bruno2013b}%
  \BibitemOpen
  \bibfield  {author} {\bibinfo {author} {\bibfnamefont {P.}~\bibnamefont
  {Bruno}},\ }\href {\doibase 10.1103/PhysRevLett.111.070402} {\bibfield
  {journal} {\bibinfo  {journal} {Phys. Rev. Lett.}\ }\textbf {\bibinfo
  {volume} {111}},\ \bibinfo {pages} {070402} (\bibinfo {year}
  {2013}{\natexlab{c}})}\BibitemShut {NoStop}%
\bibitem [{\citenamefont {Watanabe}\ and\ \citenamefont
  {Oshikawa}(2015)}]{Watanabe2015}%
  \BibitemOpen
  \bibfield  {author} {\bibinfo {author} {\bibfnamefont {H.}~\bibnamefont
  {Watanabe}}\ and\ \bibinfo {author} {\bibfnamefont {M.}~\bibnamefont
  {Oshikawa}},\ }\href {\doibase 10.1103/PhysRevLett.114.251603} {\bibfield
  {journal} {\bibinfo  {journal} {Phys. Rev. Lett.}\ }\textbf {\bibinfo
  {volume} {114}},\ \bibinfo {pages} {251603} (\bibinfo {year}
  {2015})}\BibitemShut {NoStop}%
\bibitem [{\citenamefont {Sacha}(2015{\natexlab{a}})}]{Sacha2015}%
  \BibitemOpen
  \bibfield  {author} {\bibinfo {author} {\bibfnamefont {K.}~\bibnamefont
  {Sacha}},\ }\href {\doibase 10.1103/PhysRevA.91.033617} {\bibfield  {journal}
  {\bibinfo  {journal} {Phys. Rev. A}\ }\textbf {\bibinfo {volume} {91}},\
  \bibinfo {pages} {033617} (\bibinfo {year} {2015}{\natexlab{a}})}\BibitemShut
  {NoStop}%
\bibitem [{\citenamefont {Sacha}(2015{\natexlab{b}})}]{Sacha2015a}%
  \BibitemOpen
  \bibfield  {author} {\bibinfo {author} {\bibfnamefont {K.}~\bibnamefont
  {Sacha}},\ }\href@noop {} {\bibfield  {journal} {\bibinfo  {journal} {Sci.
  Rep.}\ }\textbf {\bibinfo {volume} {5}},\ \bibinfo {pages} {10787} (\bibinfo
  {year} {2015}{\natexlab{b}})}\BibitemShut {NoStop}%
\bibitem{else16}  
  D. V. Else, B. Bauer, and C. Nayak, arXiv:1603.08001.
\bibitem [{\citenamefont {Buchleitner}\ \emph {et~al.}(2002)\citenamefont
  {Buchleitner}, \citenamefont {Delande},\ and\ \citenamefont
  {Zakrzewski}}]{Buchleitner2002}%
  \BibitemOpen
  \bibfield  {author} {\bibinfo {author} {\bibfnamefont {A.}~\bibnamefont
  {Buchleitner}}, \bibinfo {author} {\bibfnamefont {D.}~\bibnamefont
  {Delande}}, \ and\ \bibinfo {author} {\bibfnamefont {J.}~\bibnamefont
  {Zakrzewski}},\ }\href
  {http://www.sciencedirect.com/science/article/pii/S0370157302002703}
  {\bibfield  {journal} {\bibinfo  {journal} {Physics reports}\ }\textbf
  {\bibinfo {volume} {368}},\ \bibinfo {pages} {409} (\bibinfo {year}
  {2002})}\BibitemShut {NoStop}%
\bibitem{MuellerDelande:Houches:2009}  
  C. A. M\"uller and D. Delande, Disorder and interference:
  localization phenomena, in {\it Ultracold Gases
  and Quantum Information}, Lecture Notes of the Les Houches Summer School in Singapore Vol.~91, July 2009, edited by C. Miniatura {\it et al.} (Oxford University Press, Oxford, 2011), Chap.~9, arXiv:1005.0915.
\bibitem [{\citenamefont {Kuhn}\ \emph {et~al.}(2007)\citenamefont {Kuhn},
  \citenamefont {Sigwarth}, \citenamefont {Miniatura}, \citenamefont
  {Delande},\ and\ \citenamefont {M\"uller}}]{Kuhn:Speckle:NJP07}%
  \BibitemOpen
  \bibfield  {author} {\bibinfo {author} {\bibfnamefont {R.~C.}\ \bibnamefont
  {Kuhn}}, \bibinfo {author} {\bibfnamefont {O.}~\bibnamefont {Sigwarth}},
  \bibinfo {author} {\bibfnamefont {C.}~\bibnamefont {Miniatura}}, \bibinfo
  {author} {\bibfnamefont {D.}~\bibnamefont {Delande}}, \ and\ \bibinfo
  {author} {\bibfnamefont {C.~A.}\ \bibnamefont {M\"uller}},\ }\href {\doibase
  10.1088/1367-2630/9/6/161} {\bibfield  {journal} {\bibinfo  {journal} {New J.
  Phys.}\ }\textbf {\bibinfo {volume} {9}},\ \bibinfo {pages} {161} (\bibinfo
  {year} {2007})}\BibitemShut {NoStop}%
\bibitem [{Note1()}]{Note1}%
  \BibitemOpen
  \bibinfo {note} {The role of the parameter $\alpha $ is to remove the
  degeneracy between the two counter-propagating waves at $p=\pm \omega $ which
  are simultaneously resonant. If $\alpha =0$ one would deal with (however very
  small) tunneling effects between counter propagating particles and the true
  eigenstates would be linear combinations of the states considered here and
  their counter-propagating mirror images.}\BibitemShut {Stop}%
\bibitem [{\citenamefont {Lugan}\ \emph {et~al.}(2009)\citenamefont {Lugan},
  \citenamefont {Aspect}, \citenamefont {Sanchez-Palencia}, \citenamefont
  {Delande}, \citenamefont {Gr\'emaud}, \citenamefont {M\"uller},\ and\
  \citenamefont {Miniatura}}]{Lugan:Anderson1D:PRA09}%
  \BibitemOpen
  \bibfield  {author} {\bibinfo {author} {\bibfnamefont {P.}~\bibnamefont
  {Lugan}}, \bibinfo {author} {\bibfnamefont {A.}~\bibnamefont {Aspect}},
  \bibinfo {author} {\bibfnamefont {L.}~\bibnamefont {Sanchez-Palencia}},
  \bibinfo {author} {\bibfnamefont {D.}~\bibnamefont {Delande}}, \bibinfo
  {author} {\bibfnamefont {B.}~\bibnamefont {Gr\'emaud}}, \bibinfo {author}
  {\bibfnamefont {C.~A.}\ \bibnamefont {M\"uller}}, \ and\ \bibinfo {author}
  {\bibfnamefont {C.}~\bibnamefont {Miniatura}},\ }\href {\doibase
  10.1103/PhysRevA.80.023605} {\bibfield  {journal} {\bibinfo  {journal} {Phys.
  Rev. A}\ }\textbf {\bibinfo {volume} {80}},\ \bibinfo {pages} {023605}
  (\bibinfo {year} {2009})},\ \Eprint {http://arxiv.org/abs/0902.0107}
  {arXiv:0902.0107} \BibitemShut {NoStop}%
\bibitem{Zwerger:OpticalLattice:JPB03}  
  W. Zwerger, Journal of Optics B: Quantum and
  Semiclassical Optics {\bf 5}, S9 (2003).
\bibitem [{\citenamefont {Abramowitz}\ and\ \citenamefont
  {Stegun}(1964)}]{AbramovitzStegun:64}%
  \BibitemOpen
  \bibfield  {author} {\bibinfo {author} {\bibfnamefont {M.}~\bibnamefont
  {Abramowitz}}\ and\ \bibinfo {author} {\bibfnamefont {I.~A.}\ \bibnamefont
  {Stegun}},\ }\href@noop {} {\emph {\bibinfo {title} {{Handbook of
  Mathematical Functions}}}}\ (\bibinfo  {publisher} {{Dover}},\ \bibinfo
  {address} {{New York, USA}},\ \bibinfo {year} {1964})\BibitemShut {NoStop}%
\bibitem [{Note2()}]{Note2}%
  \BibitemOpen
  \bibinfo {note} {In general, there can be several resonances between the
  external driving at frequency $s\omega $ and the internal frequency $\partial
  H_0/\partial J.$ We assume $\lambda $ to be small enough for these resonances
  not to overlap.}\BibitemShut {Stop}%
\bibitem [{\citenamefont {Szriftgiser}\ \emph {et~al.}(1996)\citenamefont
  {Szriftgiser}, \citenamefont {Gu\'ery-Odelin}, \citenamefont {Arndt},\ and\
  \citenamefont {Dalibard}}]{Szriftgiser:Mirror:PRL96}%
  \BibitemOpen
  \bibfield  {author} {\bibinfo {author} {\bibfnamefont {P.}~\bibnamefont
  {Szriftgiser}}, \bibinfo {author} {\bibfnamefont {D.}~\bibnamefont
  {Gu\'ery-Odelin}}, \bibinfo {author} {\bibfnamefont {M.}~\bibnamefont
  {Arndt}}, \ and\ \bibinfo {author} {\bibfnamefont {J.}~\bibnamefont
  {Dalibard}},\ }\href {\doibase 10.1103/PhysRevLett.77.4} {\bibfield
  {journal} {\bibinfo  {journal} {Phys. Rev. Lett.}\ }\textbf {\bibinfo
  {volume} {77}},\ \bibinfo {pages} {4} (\bibinfo {year} {1996})}\BibitemShut
  {NoStop}%
\bibitem [{\citenamefont {Maeda}\ and\ \citenamefont
  {Gallagher}(2004)}]{Maeda2004}%
  \BibitemOpen
  \bibfield  {author} {\bibinfo {author} {\bibfnamefont {H.}~\bibnamefont
  {Maeda}}\ and\ \bibinfo {author} {\bibfnamefont {T.~F.}\ \bibnamefont
  {Gallagher}},\ }\href {\doibase 10.1103/PhysRevLett.92.133004} {\bibfield
  {journal} {\bibinfo  {journal} {Phys. Rev. Lett.}\ }\textbf {\bibinfo
  {volume} {92}},\ \bibinfo {pages} {133004} (\bibinfo {year}
  {2004})}\BibitemShut {NoStop}%
\bibitem [{\citenamefont {Bleszynski-Jayich}\ \emph {et~al.}(2009)\citenamefont
  {Bleszynski-Jayich}, \citenamefont {Shanks}, \citenamefont {Peaudecerf},
  \citenamefont {Ginossar}, \citenamefont {von Oppen}, \citenamefont
  {Glazman},\ and\ \citenamefont {Harris}}]{Bleszynski2009}%
  \BibitemOpen
  \bibfield  {author} {\bibinfo {author} {\bibfnamefont {A.~C.}\ \bibnamefont
  {Bleszynski-Jayich}}, \bibinfo {author} {\bibfnamefont {W.~E.}\ \bibnamefont
  {Shanks}}, \bibinfo {author} {\bibfnamefont {B.}~\bibnamefont {Peaudecerf}},
  \bibinfo {author} {\bibfnamefont {E.}~\bibnamefont {Ginossar}}, \bibinfo
  {author} {\bibfnamefont {F.}~\bibnamefont {von Oppen}}, \bibinfo {author}
  {\bibfnamefont {L.}~\bibnamefont {Glazman}}, \ and\ \bibinfo {author}
  {\bibfnamefont {J.~G.~E.}\ \bibnamefont {Harris}},\ }\href@noop {} {\bibfield
   {journal} {\bibinfo  {journal} {Science}\ }\textbf {\bibinfo {volume}
  {326}},\ \bibinfo {pages} {272} (\bibinfo {year} {2009})}\BibitemShut
  {NoStop}%
\bibitem{Lichtenberg_s}  
A. J. Lichtenberg and M. A. Lieberman, {\it Regular and Stochastic Motion}, Applied Mathematical Sciences Vol. 38, edited by F. John {\it et al.} (Springer, Berlin 1983).
\end{thebibliography}
%

\end{document}